\documentclass[%
 aip,
 amsmath,amssymb,
 reprint,%
]{revtex4-1}

\usepackage{graphicx}
\usepackage{dcolumn}
\usepackage{bm}
\usepackage[utf8]{inputenc}
\usepackage[T1]{fontenc}
\usepackage{mathptmx}
\usepackage{etoolbox}
\usepackage{amsmath}
\usepackage{hyperref}
\usepackage{amssymb}

\makeatletter
\def\@email#1#2{%
 \endgroup
 \patchcmd{\titleblock@produce}
  {\frontmatter@RRAPformat}
  {\frontmatter@RRAPformat{\produce@RRAP{*#1\href{mailto:#2}{#2}}}\frontmatter@RRAPformat}
  {}{}
}%
\makeatother
\begin{document}

\preprint{AIP/123-QED}

\title{Double pulse all-optical coherent control of ultrafast spin-reorientation in antiferromagnetic rare-earth orthoferrite} 

\author{N. E. Khokhlov}
    \email[]{nikolai.khokhlov@ru.nl}
    \affiliation{Radboud University Nijmegen, Institute for Molecules and Materials, 6525 AJ Nijmegen, The Netherlands}
\author{A. E. Dolgikh}
\affiliation{Radboud University Nijmegen, Institute for Molecules and Materials, 6525 AJ Nijmegen, The Netherlands}
\author{B. A. Ivanov}
\affiliation{Radboud University Nijmegen, Institute for Molecules and Materials, 6525 AJ Nijmegen, The Netherlands}
\affiliation{Institute of Magnetism, NAS and MES of Ukraine, 36b Vernadsky Blvd., Kiev 03142, Ukraine}
\author{A. V. Kimel}
\affiliation{Radboud University Nijmegen, Institute for Molecules and Materials, 6525 AJ Nijmegen, The Netherlands}

\date{\today}

\begin{abstract}
A pair of circularly polarized laser pulses of opposite helicities are shown to control the route of spin reorientation phase transition in rare-earth antiferromagnetic orthoferrite (Sm$_{0.55}$Tb$_{0.45}$)FeO$_3$.
The route can be efficiently controlled by the delay between the pulses and the sample temperature.
Simulations employing previously published models of \mbox{laser-induced} spin dynamics in orthoferrites failed to reproduce the experimental results.
We suggest that the failure is due to neglected temperature dependence of the antiferromagnetic resonance damping in the material.
Taking into account the experimentally deduced temperature dependence of the damping, we obtained good agreement between the simulations and the experiment.   
\end{abstract}


\maketitle 

Antiferromagnets are the largest, but probably the least explored, class of magnetically ordered materials discovered only in the 20$^{\rm{th}}$ century \cite{Landau_AFM1933, Neel_AnnPhys1948}.
The magnetic order in antiferromagnets is characterized by mutually antiparallel alignment of neighboring spins, such that their net magnetic moment is either zero or vanishingly small.
In the simplest case of a two-sublattice antiferromagnet, the order can be modeled as two ferromagnets with mutually antiparallel magnetizations of the sub-lattices \textbf{M$_1$} and \textbf{M$_2$} so that the whole material is described by the antiferromagnetic N\'eel vector $\textbf{L}=\textbf{M}_1-\textbf{M}_2$.

Due to the high frequencies of intrinsic spin resonances, often reaching the landmark of 1 THz, the antiferromagnets are seen as materials that may facilitate the fastest and least-dissipative mechanisms for writing magnetic bits in future data storage \cite{Antiferromagneticspintronics_NatNanoTech2016}.
Understanding how to control spins in antiferromagnets and revealing the characteristic time scales, which define the fundamental limits on the speed of such a control, are thus among the most heavily debated questions in contemporary magnetism \cite{HowtomanipulateAFMstates_Nanotech2018}.

Rare-earth orthoferrites have been long offering a very fruitful playground for this research.
First, because of the very strong temperature dependence of magnetic anisotropy, these materials possess a heat-induced spin-reorientation phase transition (SRT).
Thus, using the femtosecond laser pulse as an ultrafast heater, it is possible to launch spin dynamics and study spin reorientation in antiferromagnets at an unprecedentedly fast timescale \cite{kimel2004laser}.
Second, due to strong opto-magnetic effects, circularly polarized femtosecond laser pulses can act on spins in these materials as equally short pulses of effective magnetic field with the polarity defined by the helicity of light \cite{Kimel2005Nature}. 

A combination of these two mechanisms of launching the spin dynamics led, in particular, to the discovery of spin inertia in antiferromagnets \cite{Kimel_inertiaNatPhys2009} and to the routes of coherent control of SRT \cite{deJong_CoherentControl_PRL2012}.
Although intuitively heat-induced SRT can proceed along two energetically equivalent routes with the material eventually ending up in a multidomain state, ultrashort pulses of opto-magnetic fields were suggested to dynamically break the degeneracy and steer the medium to a state defined by the helicity of the light pulse \cite{deJong_CoherentControl_PRL2012}.
Later, the same principle of dynamical degeneracy breaking was employed to demonstrate coherent control of SRT in orthoferrites with the help of a pair of pulses -- a properly timed femtosecond laser heat pulse and a nearly single-cycle pulse of the THz magnetic field \cite{Kurihara_MacroscopicPRL2018}.

\begin{figure}
\includegraphics[width=1 \linewidth]{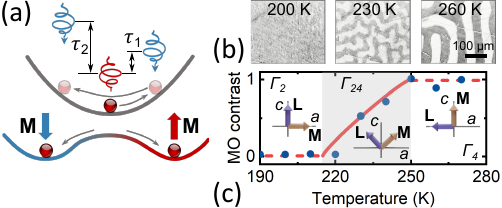}
\caption{\label{fig:idea}
(a) Schematic illustration of the coherent optical control of final magnetization state at SRT.
(b) Magneto-optical images of the sample at different temperatures in Faraday geometry.
(c) Temperature dependence of magneto-optical contrast between opposite domains (symbols).
Line is the theoretical dependence found using the model from \cite{Horner_NatureofSRT_PRL1968}.
Insets depict the orientations of antiferromagnetic vector $\mathbf{L}$ and net magnetization $\mathbf{M}$ of the sample in the low-temperature ($\Gamma_{2}$), angular ($\Gamma_{24}$), and high-temperature ($\Gamma_{4}$) phases.
}
\end{figure}

Here, we further explore the coherent control of SRT with a pair of optical pump pulses.
Employing two circularly polarized pulses acting as both ultrafast heater and opto-magnetic field, the time delay of the second-arrived pump should define the final magnetization orientation in a controllable manner [Fig.\ref{fig:idea}(a)].
In experiments, we reveal a strong and previously ignored effect of heavily increased damping of spin precession near the phase transition.
The damping significantly affects the result of the action of the pair of pulses.
We showed that if the first pulse heats the orthoferrite to a temperature near the SRT, the material becomes practically insensitive to the second pulse in the subsequent time window of 5-20 ps.
According to the simulations based on the models employed before, such an insensitivity can indeed be observed, but in this case it must be observed periodically at later time delays as well.  
We propose an upgrade for the model that accounts for the increased damping and enables a match of the modeling with the experimental results. 

Here, we employ (Sm$_{0.55}$Tb$_{0.45}$)FeO$_3$ as a sample.
The crystal was grown using the floating zone technique \cite{BALBASHOV_Apparatus_for_growth_JCG_1981}.
For the study, the bulk crystal is cut in the form of 158-$\mu$m-thin plane-parallel plate with normal along $c$ axis (Appendix \ref{appendix:xray}).
The magnetic structure of the crystal can be modeled as a two-sublattice antiferromagnet with magnetizations \textbf{M}$_1$ and \textbf{M}$_2$, respectively.
The exchange interaction favors their mutually antiparallel orientations, but due to Dzyaloshinskii-Moriya interaction \textbf{M}$_1$ and \textbf{M}$_2$ are slightly canted by about 1$^\circ$, resulting in non-zero net magnetization $\textbf{M} = \textbf{M}_1 + \textbf{M}_2 \neq 0 $.
As a result, magnetization \textbf{M} and the antiferromagnetic N\'eel vector ${\bf L}$ are orthogonal to each other [insets on Fig.\,\ref{fig:idea}(c)].
At temperatures $T <$ 215 K, the spins and \textbf{L} are aligned along $c$ crystallographic axis, while \textbf{M} is along $a$ axis ($\Gamma_{2}$ phase).
Due to the strong temperature dependence of magneto-crystalline anisotropy, in the range 215 K$ < T < $ 250 K the spins continuously rotate within $ac$ plane ($\Gamma 24$ phase).
At $T >$ 250 K, the spins are along $a$ axis, while \textbf{M} is along $c$ axis ($\Gamma 4$ phase).
We experimentally confirmed the presence of SRT in the sample, measuring the Faraday effect for light propagating along $c$ axis and therefore sensitive to out-of-plane magnetization component [Fig.\,\ref{fig:idea}(b,c)]. 
The increase in temperature promotes spin reorientation from phase $\Gamma 2$ with in-plane \textbf{M} orientation and indistinguishable domain structure [left panel in Fig.\,\ref{fig:idea}(b)] to phase $\Gamma 4$ along one of the two equivalent routes leading either to a state with magnetization "up" or to a state with magnetization "down", appearing as bright and dark domains on camera [central and right panels in Fig.\,\ref{fig:idea}(b)].
In addition, we performed complementary dilatometry \cite{Kuchler_dilatometer2012} in zero field and the same temperature range to observe spin reorientation more clearly.
Unfortunately, the orthoferrite sample does not reveal a detectable magnetostriction signal.

The idea of our experiment is to control the route with a pair of laser pulses and reveal how the final state depends on the time delay between the pulses in a pair.
An expected scenario that we aim to verify is shown in Fig.\,\ref{fig:idea}(a).
The first pulse acts as an ultrafast heater and a pulse of opto-magnetic field.
Hence, it launches a low-amplitude spin precession and simultaneously causes transient changes in thermodynamic equilibrium.
Using a properly timed second pulse, which also acts as an ultrafast heater and an ultrashort pulse of the opto-magnetic field, one can push the spin system either to the state with the magnetization "up" or "down". 

To study the magnetization reorientation induced by such a double-pulse excitation, we used a time-resolved magneto-optical pump-probe technique combined with magneto-optical imaging \cite{Vahaplar_UltrafastPath_PRL2009}.
The sample is pumped with two 50 fs circularly polarized laser pulses with a central wavelength of 800 nm, generated by Ti:sapphire amplifier at 1 kHz repetition rate.
The time delay between the two pumps $\tau$ is mechanically controlled in a range from -150 to +150 ps.
The pulses follow the same path and pump the sample at an incidence angle of 11$^\circ$.
The focus spot on the sample has full width at a half maximum of 100 $\mu$m.
Pump-induced changes in the sample are probed with a linearly polarized pulse with a wavelength converted from 800 to 650 nm using an optical parametric amplifier.
The probe is unfocused to cover an area of about 3\,mm$^2$ on the sample with a fluence four orders of magnitude lower than that of each pump pulse.
The time delay between the first-arrived pump and the probe is mechanically controlled from -0.5 to +1.5\,ns.
Two complementary sets of experiments are performed.
The first used a CCD camera as a detector to obtain magneto-optical images of the sample \cite{Hashimoto_MOimagingRevSciInstr2014}.
In the second set of experiments, a diaphragm is placed in the probe beam, selecting only the pumped area.
After this spatial filtering, the probe is detected with a balanced detector and lock-in amplifier, synchronized with a mechanical chopper placed on the pumps' path.
In both cases, the measurements are sensitive to the out-of-plane component of magnetization parallel to $c$ axis.
The sample is placed in a cold finger cryostat to control its initial temperature $T$. 
The experiments were performed without an external magnetic field.
The helicities of the two pumps are set to be right-handed  $\sigma^+$ and left-handed $\sigma^-$ with quarter-wave plates.

\begin{figure}
\includegraphics[width=1\linewidth]{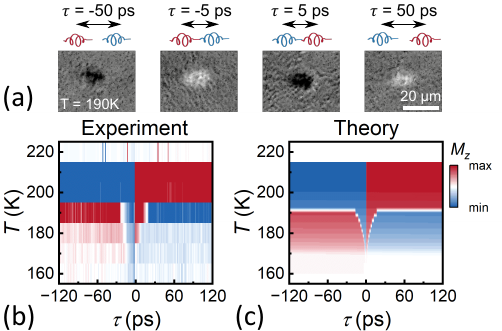}
\caption{\label{fig:DoublePumpSwitch}
(a) Magneto-optical images of the reorientation process at $t=1.5$\,ns for different pump-pump delays $\tau$ observed with CCD camera at $T = 190$\,K.
(b) Experimental diagram of magnetization's final state in coordinates of pump-pump delay $\tau$ and initial temperature of the sample $T$.
(c) The diagram of the final state calculated with Eq.\,\eqref{eq:oscillator}.
The color code shows the value $\cos{\theta}$, as the experimental scheme is sensitive to out-of-plane component of magnetization.
}
\end{figure}

For double-pump experiments, we set the fluences of the pump pulses to 67 mJ/cm$^2$ to such that one pump alone could not launch the SRT at $T=170$\,K, but two pump pulses together were able to initiate and steer the phase transition.
Figure \ref{fig:DoublePumpSwitch}(a) shows that the sign of magnetization in the final state, measured at 1.5 ns after laser excitation, dramatically depends on the time delay between pump pulses $\tau$.

To obtain a better understanding of the observed dynamics, we performed double-pump experiments with the balanced photodetector as a function of temperature $T$ and time delay $\tau$.
In Fig.\,\ref{fig:DoublePumpSwitch}(b) we plot experimentally defined diagram showing the magneto-optical contrast of the pumped domain at 1.5 ns after the laser excitation as functions of $T$ and $\tau$. 
At temperatures just below the SRT, i.e. $190 < T < 215$\,K, the double-pump excitation forms a single domain with the magnetization direction defined by the helicity of the earliest pump pulse independently on the pump-pump time delay $\tau$.
At lower temperatures $170 \leq T \leq 190$\,K, the delay $\tau$ begins to play a crucial role.
If $\tau$ is less than a critical value $\tau_c$, the helicity of the first-arrived pump determines the final orientation of magnetization.
However, if $\tau > \tau_c$, the magnetization orientation is defined by the helicity of the latest pulse.
At lower temperatures of $T < 170$\,K, two pumps together do not sufficiently heat the system to trigger SRT.

To simulate laser-induced spin dynamics, we solved the equation of motion for the antiferromagnetic vector \textbf{L} derived using the principles of Lagrangian mechanics~\cite{Zvezdin_DynamicsDW_1979, Baryakhtar_DynamicsDWSPhU1985}.
The resulting dynamics of \textbf{L} within $ac$ plane is described by the angle $\theta$ between \textbf{L} and $a$ axis as~\cite{Ivanov_Spindynamics_LTP2014, Kimel_inertiaNatPhys2009}:

\begin{equation}\label{eq:oscillator}
    \frac{d^2\theta}{dt^2} + 2\zeta\frac{d\theta}{dt} + \gamma H_{ex} \frac{dW_a(\theta)}{d\theta} = \gamma^2 H_D H_{p}(t)\sin{\theta},
\end{equation}
where the total length of \textbf{L} is assumed to be conserved;
$\zeta$ is a damping parameter in the units of frequency;
$H_p(t)$ is the pulse effective opto-magnetic field with duration of 50 fs and aligned either parallel ($\sigma^+$) or antiparallel ($\sigma^-$) with respect to $c$ axis;
$H_D$ is the effective field of the Dzyaloshinskii-Moriya interaction;
$H_{ex}$ represents the exchange field of the antiferromagnet;
$\gamma$ is the gyromagnetic ratio;
$t$ is time after the earliest pump.
The function $W_a(\theta)$ is the potential energy described by the magnetic anisotropy of the antiferromagnet.
In our model, $W_a(\theta)$ is a function of temperature in accordance with the conventionally accepted model~\cite{Horner_NatureofSRT_PRL1968}.
To mimic laser-induced heating, we assume that the temperature of the sample is a function of the pump-probe time delay, similarly to the model from Ref.~\cite{deJong_dynamicsErFeO3PRB2011}.
In particular, we take into account that the time dependence of magnetic anisotropy is due to temperature-induced repopulation of the electronic states in highly anisotropic Sm$^{3+}$ and Tb$^{3+}$ ions.
This repopulation occurs on a time scale of electron-phonon interaction for rare-earth ions, which could be estimated to be around 15 ps \cite{deJong_dynamicsErFeO3PRB2011}. 
Furthermore, we suggest an increase in temperature of 25 K after one
pump, as the action of both pumps is enough to induce SRT at $T = 170$\,K.  
The results of the modeling are similar to those from Ref. \cite{Kurihara_MacroscopicPRL2018} (see Appendix \ref{appendix:lowDamping}), but they are clearly different with respect to the experimental observations.
In the modeling, we indeed observe a triangle centered around $\tau=0$, similar to the experimental diagram in Fig. \ref{fig:DoublePumpSwitch}(b).
This triangle reproduces the insensitivity of spins in the antiferromagnet to the second pump pulse.
However, contrary to the experiment, this insensitivity also appears periodically at longer $\tau$ in the simulations (Appendix \ref{appendix:lowDamping}), but is clearly absent in the experiment [Fig. \ref{fig:DoublePumpSwitch}(b)]. 
However, we note that none of the models suggested before took into account the fact that the damping parameter $\zeta$ in Eq.\,\eqref{eq:oscillator} must also have a strong temperature dependence.
The opto-magnetic pulse triggers spin oscillations at the frequency of the quasi-ferromagnetic mode of the antiferromagnetic resonance in the orthoferrite \cite{Kimel2005Nature}.
This mode is known to "soften" down to zero frequency at temperatures of SRT from $\Gamma_{2}$ to $\Gamma_{24}$, as well as from $\Gamma_{24}$ to $\Gamma_{4}$.
It is a well-known experimental fact that softening of magnetic resonances is accompanied by a dramatic increase in damping.
This is also the case in our experiment.
Indeed, we found the damping peak at $T_1$ in single-pump experiments, extracting the frequency and damping of the spin oscillations as a function of temperature (Appendix \ref{appendix:singlePump}).
If we add this experimentally defined dependence $\zeta(T)$ to Eq.\,\eqref{eq:oscillator}, the results of the simulations appear to be in good qualitative agreement with the experiment [Fig.\ref{fig:DoublePumpSwitch}(c)].  

\begin{figure}
\includegraphics[width=1\linewidth]{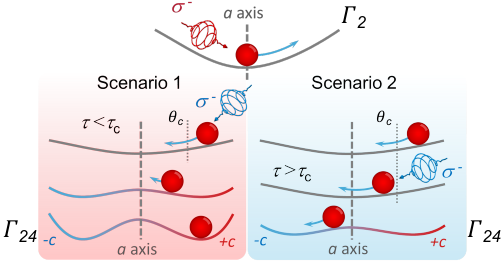}
\caption{\label{fig:Theory}
Two scenarios of the coherent magnetization control under double-pulse fluence.
The final state is defined by pumps' helicities and pump-pump delay $\tau$.
Detailed description is in the text.
}
\end{figure}

On the basis of the experimental observations and modeling, we suggest the next path of the reorientation process under double-pump excitation. 
In phase $\Gamma_{2}$ ($T<T_1$) the equilibrium orientation of \textbf{M}-\textbf{L} pair corresponds to $\theta_0 = \pm \pi/2$.
To be specific in the illustration, we assume the initial state with $\theta_0 = \pi/2$, and the first-arriving pump with positive helisity $\sigma^+$ launches the magnetization dynamics through IFE towards $\theta > \theta_0$ (Fig.\,\ref{fig:Theory}, top frame).
Subsequently, there are two scenarios for the dynamics determined by $\tau$.
The first scenario is realized if the second pump arrives too early, that is, $\tau$ is smaller than a critical value $\tau_c$ (Fig.\,\ref{fig:Theory}, left frames).
In this scenario, the second pulse cannot reverse the forward motion of the system or sufficiently accelerate it in the backward motion to overcome the potential barrier in phase $\Gamma_{24}$ before the barrier appears.
Thus, the second pump will further heat the system and thus help establish the state determined by the helicity of the first pump if $\tau < \tau_c$.
The second scenario occurs if $\tau > \tau_c$, that is, the second pump arrives when the system has reversed its momentum and passed a critical coordinate $\theta_{c}$ in backward motion (Fig.\,\ref{fig:Theory}, right frames).
In this scenario, the torque of the second pulse is sufficient to transfer the system to $\theta < \theta_0$ before the potential barrier appears in $\Gamma_{24}$.
We notice that the same scenarios work for the initial combination of $\sigma^+$ and $\theta_0 = -\pi/2$, since the torque induced by IFE does not change sign with the sign of $\theta_0$ \cite{Hansteen_NonthermalultrafastPRB2006, Shelukhin_YIGPRB2018}.
Thus, helicity $\sigma^+$ works in the same way for both initial orientations of the domains in $\Gamma_{2}$.
The change in helicity sign flips the initial torque and the corresponding final state, resulting in the antisymmetrical $T-\tau$ diagram with respect to $\tau = 0$.

In conclusion, we experimentally and numerically studied coherent control of the ultrafast phase transition in antiferromagnetic rare-earth orthoferrite using double-pulse excitation.
We show that the final state at 1.5 ns after pump excitation depends on the time delay between the pump pulses.
At a temperature close to the phase transition, the final state is fully defined by the helicity of the earliest pump pulse.
At lower temperature, we distinguish two regions.
In particular, we show that at pump-pump delays larger than a critical time, the state is defined by the helicity of the latest pulse, while at shorter delays, it is the earliest pulse in the pair that defines the final magnetization.
We show that earlier published models are unable to reproduce the experimental results and suggest that the reason for the discrepancy is the neglected temperature dependence of the damping.
Finally, we note that the peak-like behavior of damping at SRT temperature is more general and inherent to other kinds of phase transitions, where the softening of the corresponding mode appears: damping peak of magnons at antiferromagnetic-paramagnetic phase transition at N\'eel temperature \cite{bossini2016macrospin, Afanasiev_ScienceAdvances2021, Zhang_FePS3NanoLett2021}; easy-axis to easy-plane Morin transition in antiferromagnets \cite{lebrun_long_Nature2020}; damping peak of the mode with out-of-phase magnetizations precession in synthetic antiferromagnets at spin-flop transition \cite{Sorokin_syntheticantiferromagnetsPRB2020}; decrement of spin-lattice relaxation time in nuclear quadrupole resonance studies of structural phase transitions in cubic antifluorite and cubic perovskite structures \cite{armstrong1975pure}; peak of the relative absorption coefficient of dynamics in multiferroic crystals at magnetic and ferroelectric phase transitions \cite{golovenchits2004magnetic}.

\section*{Acknowledgments}
We thank P. Tinnemans for XRD measurements; S. Wiedmann and T. Ottenbros for dilatometry measurements; T. Blank and K. Saeedi Ilkhchy for fruitful discussions.
The work is supported by the European Research Council ERC Grant Agreement No. 101054664 (SPARTACUS).

\section*{Conflict of Interest Statement}
The authors have no conflicts to disclose.

\section*{Author Contributions}
\textbf{Nikolai E. Khokhlov}: Investigation (equal); Methodology (experiment); Visualization (equal); Data Curation; Writing -– original draft (equal); Writing –- review and editing (equal).
\textbf{Alexander E. Dolgikh}: Investigation (equal); Software; Visualization (equal); Writing -– original draft (equal).
\textbf{Boris A. Ivanov}: Methodology (theoretical model); Writing –- original draft (equal).
\textbf{Alexei V. Kimel}: Conceptualization; Supervision; Writing –- review and editing (equal); Project Administration.

\section*{Data availability statement}
The source data of the figures are publicly available at \url{https://doi.org/10.34973/5nxv-2q76}.
All other data supporting the findings of this article are available from the corresponding author upon request.

\appendix

\section{Single-crystal x-ray diffraction}\label{appendix:xray}

Single-crystal x-ray diffraction (XRD) confirms the sample under the study is $c-$cut single crystal with orthorhombic $P$ structure.
The crystal lattice parameters are $a=  5.38$\,\AA, \mbox{$b=  5.58$\,\AA,} $c=  7.60$\,\AA.

\setcounter{figure}{0}
\renewcommand \thefigure {A\arabic{figure}}
\begin{figure}[h]
\includegraphics[width=1\linewidth]{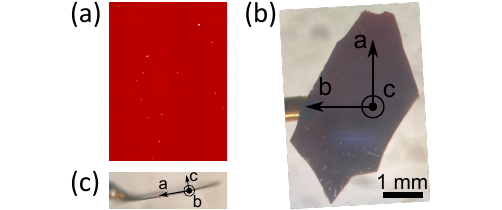}
\caption{\label{fig:Xray}
(a) Example of the measured XRD pattern.
(b,c) Photos of the sample with orientation of the crystallographic axes.
} 
\end{figure}

\section{Simulations with low damping parameter}\label{appendix:lowDamping}

\begin{figure}[h]
\includegraphics[width=1\linewidth]{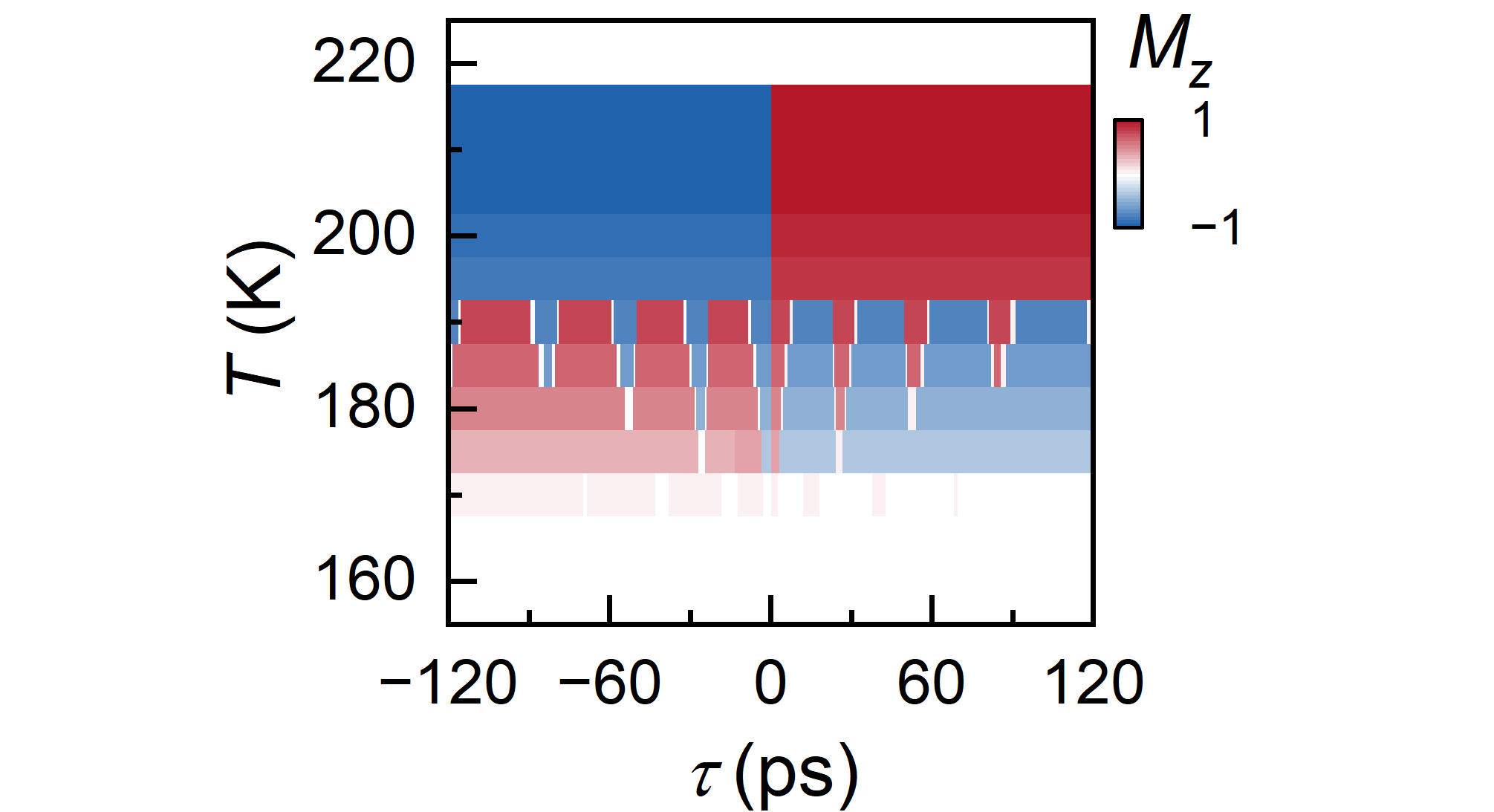}
\caption{\label{fig:AppendixDiagram}
Diagram of the final state of magnetization in the coordinates of the pump-pump delay $\tau$ and the initial temperature of the sample $T$, calculated with Eq.\,\eqref{eq:oscillator} at fixed dimensionless damping parameter \textbf{\textit{$\zeta/(2\pi f) = 0.1$}}, obtained in single-pump experiments at $T=170$\,K (see Fig.\,\ref{fig:app_sine}). 
} 
\end{figure}

\section{Single-pump experiments at low pump fluence}\label{appendix:singlePump}

\begin{figure}
\includegraphics[width=1\linewidth]{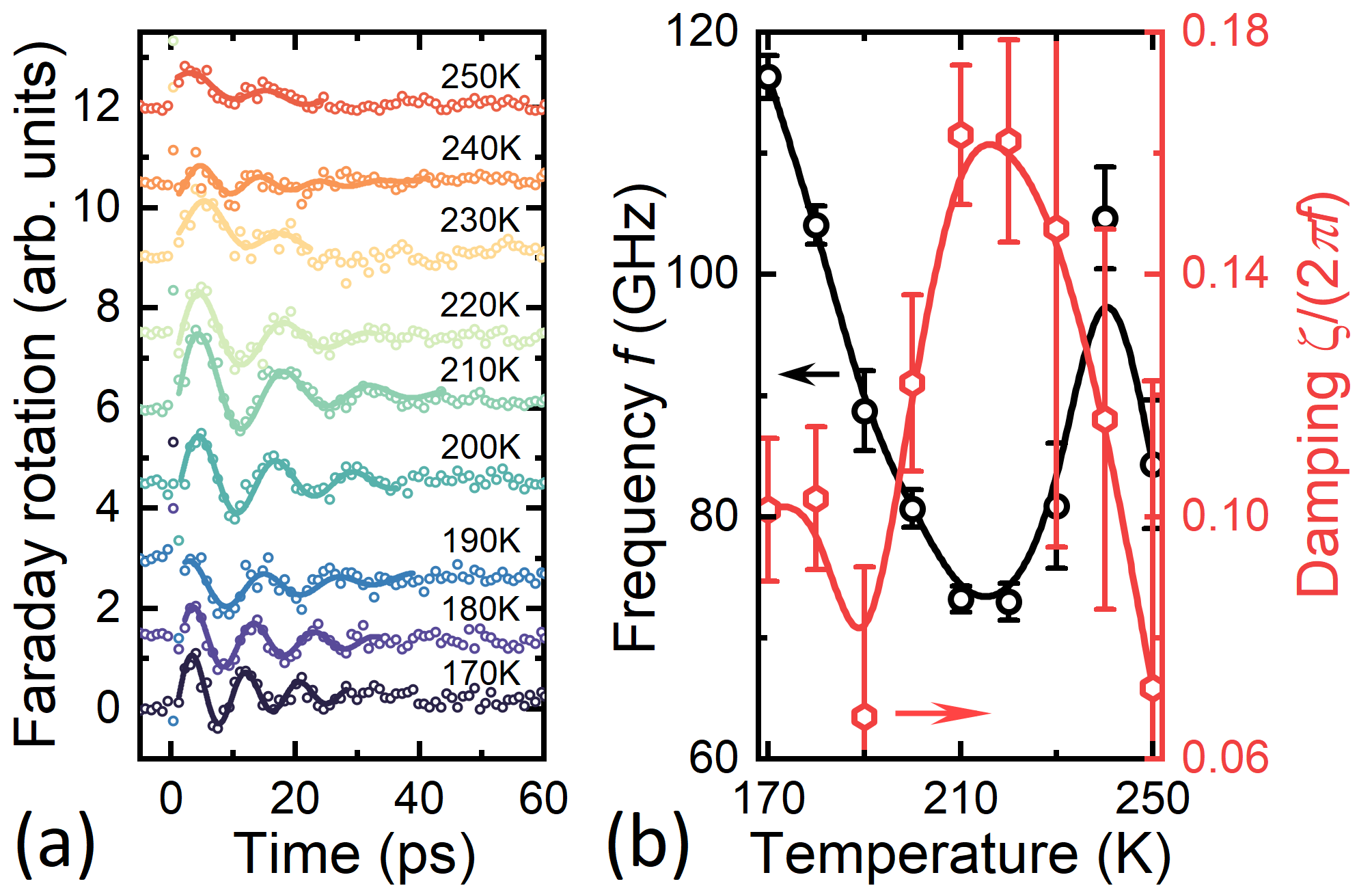}
\caption{\label{fig:app_sine}
(a) Magnetization dynamics in single-pump experiments at different initial temperatures. 
The difference of two pump-probe signals at opposite pump helicities $\sigma^+$ and $\sigma^-$ is shown.
Symbols -– experiment; lines –- fits with damping sine function.
The data sets are shifted along the vertical axis for the convenience.
(b) Temperature variation of the frequency $f$ and the dimensionless decay rate $\lambda = \zeta/(2\pi f)$, estimated from data in panel (a).
Solid lines are guides to the eye.
} 
\end{figure}

Figure \ref{fig:app_sine} represents experimental data on single-pump excitation with low fluence of 25 mJ/cm$^2$.
The oscillations on Fig.\,\ref{fig:app_sine}(a) are fitted with the function
\mbox{$F(t) = A \exp{(-2\pi f \lambda t)} \sin(2\pi f t + \varphi) + y(t)$}, where $F$ is Faraday signal, $\lambda = \zeta/(2\pi f)$ is the dimensionless decay rate; $f, A, \varphi$ are the frequency, amplitude, and initial phase of oscillations, respectively; $y(t)$ is a slow varying offset.

\section*{References}
\bibliography{DP_bibliography}

\end{document}